# Molecular dynamics study on the microstructure of $CH_3COOLi$ solutions with different concentrations

Guoyu Tan, Jiaxin Zheng* and Feng Pan*

*School of Advanced Materials, Peking University, Shenzhen Graduate School, Shenzhen, 518055, People's Republic of China*

Corresponding Authors. E-mail: panfeng@pkusz.edu.cn, zhengjx@pkusz.edu.cn.

Due to the toxic and flammable problems of organic electrolytes, the study on concentrated aqueous system for lithium ion batteries (LIBs) has attracted wide attention. In this paper, by molecular dynamics simulations, the $CH_3COOLi$ aqueous system is considered as the potential concentrated aqueous system for LIBs, and the all variations of the microstructure of the aqueous system from dilution to concentration are analyzed. The details of microstructure are discussed, especially the interactions concerning anions. Among them, the first peak of RDF (radial distribution function) between the $Li^+$ ion and the oxygen atom in $CH_3COOLi$ is 2.9 Å, which does not change from dilution to concentration. This RDF information further indicates that when the concentration increases, the microstructures of small components formed by any two clusters do not change much, but at same time, the spatial structures constructed by many small components are gradually built up from a broader perspective.

The traditional lithium ion batteries (LIBs) have achieved great success in commercial applications[1-5]. However, with the increasing demand for energy storage quality, the disadvantages of traditional lithium-ion batteries are also becoming more and more apparent. How to get over the flammability of traditional organic electrolytes has become an important research topic[6]. Taking aqueous Li-ion battery as an alternative is one of the most popular and essential approaches[4, 7]. In overall respects, aqueous electrolytes can get high scores, but the only drawback is its narrow electrochemical window, since under the classical system, the maximum electrochemical window that can be obtained in aqueous full Li-ion batteries is 1.5 V[8-10]. It has been a scientific challenge to eliminate this ultimate restriction by stabilizing water thermodynamically, especially on the anode side where water is reduced so as to evolve hydrogen. It seems evident that we can change the electrochemical window by changing the pH value, but in fact, the pH value changes the voltage of each electrode while the voltage difference between the two electrodes is constant[8]. In addition, there is also an attempt to create an interface similar to SEI on the surface of the electrode, and in so doing, it is expected to obtain a larger electrochemical window with the aid of the interface. But for a long time, the formation of a dense SEI is considered as a very difficult mission, given the simple and special elements of the water molecules[11-12].

It is not until the discovery of organic electrolytes that when the concentration of solute is very high, the physical and chemical properties of the solutions have many abnormal changes, such as extremely strong thermodynamic stability, the stability of redox reaction and so on, which provides a new direction for the exploration of the problem of aqueous electrolytes[13]. Recently, Suo *et al.* conducted the super concentration research in the aqueous electrolyte for the first time and found that the electrochemical window of the aqueous electrolyte was expanded[7]. LiTFSI is used as lithium salt in the aqueous system with a concentration of 21 mol/L (equivalent to the proportion of $Li^+$: $H_2O$ = 1:2.61), reaching an electrochemical window close to 3.0 V. Then they make a full battery with a voltage of 2.3 V and suggest two possible reasons to explain this phenomenon: one is the destruction of the local structure of lithium ions by the anions; the second is that the expanding of the electrochemical window may arise from the formation of LiF on the negative electrode surface. Subsequently, Yamada *et al.* obtained a more concentrated aqueous electrolyte system mixed with two kinds of lithium salts including LiTFSI, and in this way, the electrochemical window reaches 3.1 V[14].

They consider that in the super concentrated aqueous solutions, there are whole new structures appeared and one of the features is that free water molecules not coordinated with $Li^+$ ions are disappeared. This special microstructure of the super concentrated solutions determines its properties including electrochemical window, and thus it is very important to study the microstructure of the super concentrated solutions. It is well-known that in the aqueous solutions, the distance between $Li^+$ ion and water molecule is 2.1 Å and invariable in all kinds of concentrations[15-21]. However, anions, which also play an important role in the formation of the spatial construction, are rarely focused on. It has been reported that $CH_3COOLi$ has a crystal structure of $CH_3COOLi \cdot 2H_2O$, and at normal temperature, it has a melting state close to $Li^+$: $H_2O$ = 1:2[22-23], so $CH_3COOLi$ can be proposed as a good candidate as solute in aqueous electrolyte.

In this work, anions are paid special attention on and $CH_3COOLi$ is selected as the research object. Molecular dynamics is used to study the spatial micro configuration of the $CH_3COOLi$ aqueous electrolyte system from dilution to concentration. First, the micro configurations of six concentrations ($Li^+$: $H_2O$ = 1:32, $Li^+$: $H_2O$ = 1:8, $Li^+$: $H_2O$ = 1:5.3, $Li^+$: $H_2O$ = 1:4, $Li^+$: $H_2O$ = 1:3.5, $Li^+$: $H_2O$ = 1:3) have been fully described. Then we compare the main interactions concerning anions. Based on the information provided by RDF (radial distribution function), when the concentrations increase, the biggest variation is the probability of the occurrence of each RDF peak, whereas the RDF peaks are constants. This indicates that the microstructures of small components formed by any two clusters do not change much, but at same time, the spatial structures constructed by many small components are gradually built up from a broader perspective.

**Molecular dynamics simulations methodology.** Molecular dynamics (MD) simulations were performed to study the variations of micro configurations of $CH_3COOLi$ aqueous electrolyte system. The MD simulations were run using LAMMPS package[24-27]. The tip3p force field was used to model interactions between the $H_2O$[28]. In all these instances the interaction energy is given by[29-31],

$$V_{(r)} = \sum_{bonds} k_b(r - r_0)^2 + \sum_{angles} k_\theta(\theta - \theta_0)^2 \\ + \sum_{dihedrals} k_\chi[1 + \cos(n_0 - \delta_0)] \\ + \sum_{impropers} k_\psi(\psi - \psi_0)^2 \\ + \sum_{i=1}^{N-1} \sum_{j=i+1}^{N} \left\{ 4\varepsilon_{ij} \left[ \left(\frac{\sigma_{ij}}{r_{ij}}\right)^{12} - \left(\frac{\sigma_{ij}}{r_{ij}}\right)^6 \right] \\ + \frac{q_i q_j}{r_{ij}} \right\}$$

where the symbols represent their conventional meaning. Lennard-Jones (LJ) interaction parameters were used with a cutoff distance of 12 Å. Periodic boundary conditions were applied in the three directions. All the parameters for the $CH_3COO^-$ anion used in this work are summarized in **Table 1**.

**Table 1**. Force field Parameters Used in This Work for the CH3COO- Anion

| Atom type | Charge (e) | σ (Å) | ε (kcal·mol$^{-1}$) |
|---|---|---|---|
| $H_O$ | 0.41 | 0 | 0 |
| $O_H$ | -0.82 | 3.15 | 0.15207 |
| $O_C$ | -0.861 | 2.96 | 0.20923 |
| $C_O$ | 0.972 | 3.75 | 0.10499 |
| $C_H$ | -0.25 | 3.50 | 0.65759 |
| $H_C$ | 0.0 | 2.42 | 0.01499 |

| Bond type | Force Constant k$_b$ (kcal·mol-1·Å-2) | r0(Å) |
|---|---|---|
| $H_O$—$O_H$ | 540.6 | 0.957 |
| $C_O$—$C_H$ | 247.5 | 1.542 |
| $C_O$—$O_C$ | 619.0 | 1.256 |
| $C_H$—$H_C$ | 373.0 | 1.089 |

| Angle type | Force Constant k$_\theta$ (kcal·mol$^{-1}$·rad$^{-2}$) | θ$_0$(deg) |
|---|---|---|
| $H_O$—$O_H$—$H_O$ | 50.00 | 104.52 |
| $O_C$—$C_O$—$O_C$ | 131.0 | 128.83 |
| $O_C$—$C_O$—$C_H$ | 106.5 | 115.59 |
| $C_O$—$C_H$—$H_C$ | 45.5 | 110.02 |
| $H_C$—$C_H$—$H_C$ | 45.0 | 108.90 |

| Dihedral type | Force Constant k$_\chi$ (kcal·mol$^{-1}$) | n$_0$ | θ$_0$ (deg) |
|---|---|---|---|
| $H_C$—$C_H$—$C_O$—$O_C$ | 0.0089 | 6 | 0.0 |

In order to ensure that the cluster motion in the process of simulation is close to the one in the aqueous electrolyte system of the real world, it is essential to have a good equilibrium phase at the beginning. In the first step, the configuration of the initial state is based on a low dense system. It ensures that each atom has enough space to move so that the system can be close to the lowest state of the global energy, rather than the metastable state. The second step is to carry out NVT simulation with low density and higher temperature, intended to make the distribution of the system more uniform. The test indicates that setting temperature at 370K is relatively a good choice, because when the temperature is higher than 100 ℃ the temperature will make the dilute solutions boil. After verifying the uniformity of its distribution through RDF function, we can conclude that it has reached a more suitable state. The third step is to continue NPT simulation at high temperature, to fill the gaps that may exist in the system. When the volume of the system has reached a stable value and lasted for a period of time, it can be considered that the target state has been achieved. The fourth step is to use NPT to cool the system down and reach the low temperature of the target, so that the system can meet the test condition for the first time. In this step, the important criterion is that there is no mutation in the volume of the system, because the volume mutation suggests that the cooling rate is too fast. The fifth step is to simulate the NPT in the target temperature to make the system stable. After we use volume and RDF test to confirm that the state of the system is stable, we can start the formal simulation.

**Results and discussion.**
**Overview of the solutions.** By repeating the process of obtaining the equilibrium state as we mentioned before, we got 20 equilibrium systems, which are summarized in **Table 2**. A Snapshot of microstructure for $CH_3COOLi$ aqueous solutions obtained by molecular dynamics simulations after 200 ns is shown in **Fig. 1**. It can be found that, in dilute solutions, such as $Li^+$: $H_2O$ = 1:32 and 1:8, the $Li^+$ ions coordinated with water molecules freely diffuse. The water molecules not coordinated with cations and anions dominate the system, and the anions are heavier than other clusters, which however has little influence on the whole system. Cations and anions play a supportive role in this state. As a result, dilute solutions have similar physical and chemical properties with water. On the contrary, in concentrated solutions, anions, cations and water molecules are complicatedly mixed together, where new aggregations are built up. Various interactions make the solution possible to have new properties. In this respect, concentrated $CH_3COOLi$ aqueous solutions have some similarities with the concentrated LiTFSI aqueous electrolytes. This will be discussed in detail in what follows.

**Table 2**. Models of $H_2O$-$CH_3COOLi$ electrolytes for molecular dynamics (MD) simulations

| Li:$H_2O$ | Volume/ $Å^3$ | Number of atom per box | Number of $CH_3COOLi$ per box | Number of solvent per box |
|---|---|---|---|---|
| 1:32 | 24877 | 2808 | 27 | 864 |
| 1:8 | 32463 | 3456 | 108 | 864 |
| 1:5.3 | 38155 | 3888 | 162 | 864 |
| 1:4 | 87246 | 8640 | 432 | 1728 |
| 1:3.875 | 85530 | 8478 | 432 | 1674 |
| 1:3.75 | 84784 | 8316 | 432 | 1620 |
| 1:3.625 | 83179 | 8154 | 432 | 1566 |
| 1:3.5 | 81955 | 7992 | 432 | 1512 |
| 1:3.375 | 80712 | 7830 | 432 | 1458 |
| 1:3.25 | 79197 | 7668 | 432 | 1404 |
| 1:3.125 | 77653 | 7506 | 432 | 1350 |
| 1:3 | 76755 | 7344 | 432 | 1296 |
| 1:2.9 | 93640 | 9018 | 540 | 1566 |
| 1:2.8 | 92690 | 8856 | 540 | 1512 |
| 1:2.7 | 91463 | 8694 | 540 | 1458 |
| 1:2.6 | 90066 | 8532 | 540 | 1404 |
| 1:2.5 | 88770 | 8370 | 540 | 1350 |
| 1:2.4 | 87371 | 8208 | 540 | 1296 |
| 1:2.3 | 85525 | 8046 | 540 | 1242 |
| 1:2.2 | 85126 | 7884 | 540 | 1188 |

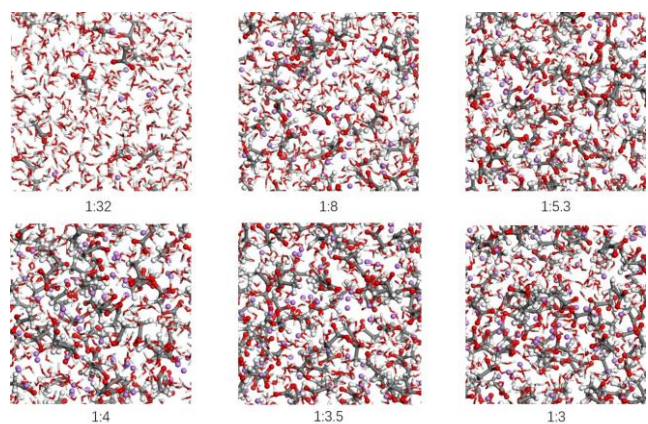

**Fig. 1.** Snapshots of microstructure for $CH_3COOLi$ aqueous solutions with different concentrations obtained by molecular dynamics simulations after 200 ns..

**RDF analyses in each aqueous electrolyte.** RDF refers to the probability of the distribution of other clusters in space to the coordinates of a given cluster. It can thus be used to characterize the micro configuration of the system. As shown in **Fig. 2**, this work selects 6 systems as representatives, and compares 8 main RDF curves within their respective systems. These 6 systems are CH$_3$COOLi aqueous solutions with concentrations of Li$^+$: H$_2$O = 1:32, 1:8, 1:5.3, 1:4, 1:3.5, 1:3. The 8 main RDF curves are Li-Li, Li-O$_H$, Li-C$_H$, O-C$_O$, O-H$_c$, H-O$_c$, H-C$_H$ and O$_c$-H$_c$, respectively. O$_H$ represents O atom in H$_2$O, C$_H$ represents C atom connected with H atom in CH$_3$COO$^-$, C$_O$ represents C atom connected with O atom in CH$_3$COO$^-$, Hc represents H atom connected with C atom in CH$_3$COO$^-$, and Oc represents O atom connected with C atom in CH$_3$COO$^-$.

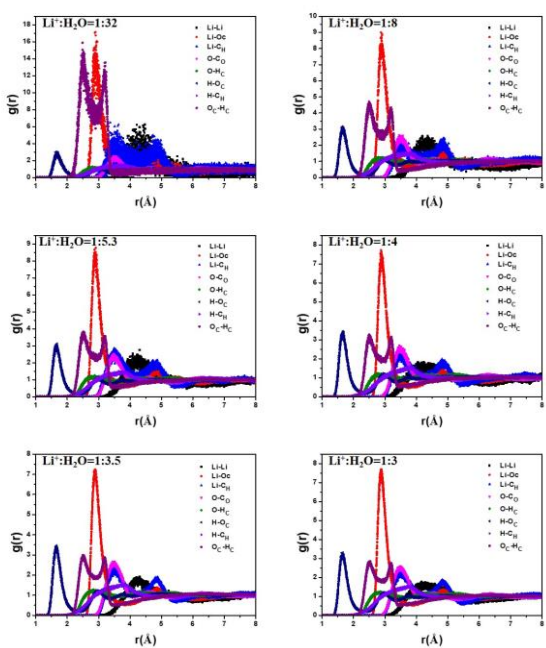

**Fig. 2.** Eight RDF curves compared in CH$_3$COOLi aqueous solutions with concentrations of Li$^+$:H$_2$O = 1:32, 1:8, 1:5.3, 1:4, 1:3.5 and 1:3. Those eight RDF curves are Li-Li, Li-O$_H$, Li-C$_H$, O-C$_O$, O-H$_C$, H-O$_C$, H-C$_H$ and O$_C$-H$_C$, respectively. O$_H$ represents O atom in H$_2$O, C$_H$ represents C atom connected with H atom in CH$_3$COO$^-$, C$_O$ represents C atom connected with O atom in CH$_3$COO$^-$, H$_C$ represents H atom connected with C atom in CH$_3$COO$^-$, O$_C$ represents O atom connected with C atom in CH$_3$COO$^-$..

When the solution is dilute, such as Li$^+$: H$_2$O = 1:32, Li-Oc is most prominent and its first peak is 2.9 Å, which suggests that the oxygen atom in the CH$_3$COO$^-$ has a strong attraction to Li$^+$ and a balance is reached near the 2.9 Å. Another more prominent peak arises from Oc-Hc interaction, which is the obvious attraction between the negative and positive ends of the anion CH$_3$COO$^-$. Besides, this peak can also be spotted in lowly concentrated solutions. At this concentration, the interactions between other atoms have not yet been highlighted. It is particularly noteworthy that the peaks of O-Co, O-Hc and H-C$_H$, which represent the interaction among water molecules and the CH$_3$COO$^-$ anion, are extremely not obvious, whereas the only detectable peak is that of H-Oc's, which, though has a certain aggregation in the vicinity of 1.7 Å, is still indistinct.

At the concentrations of Li$^+$: H$_2$O = 1:8, an obvious peak of Li-Oc is still noticeable, indicating the anion CH$_3$COO$^-$ has a strong attraction to the cationic Li$^+$. This force is the most apparent one among all 8 interaction forces under discussion. Next is the Oc-Hc interaction followed by the H-Oc interaction. These 3 interactions are all obvious. The weakest peak is created by the O-Hc interaction, indicating that the force between the water molecules and the anion CH$_3$COO$^-$ is not strong, even if the hydrogen bond interaction may be established. By contrast, the H-Oc interaction creates an obvious peak at 1.7 Å. This, similar to the hydrogen bond, is the shortest effect between the clusters in the whole system.

At the concentration of Li$^+$: H$_2$O = 1:5.3, unlike the Li-Oc interaction, the Oc-Hc interaction and the H-Oc interaction can be spotted in the dilute solutions. In addition, there are two other noticeable curves, one from the O-Co interaction, the other from the Li-C$_H$ interaction, and they all appear in the vicinity of 3.5 Å. This indicates that the C atoms in the CH$_3$COO$^-$ anion begin to participate in the interaction with the external clusters, among which the interaction between the Co atom and the O atom in the water molecules and the interaction between the C$_H$ atom and the Li$^+$ ions are especially obvious. Those interactions suggest the original electrical property in the CH$_3$COO$^-$.

When the salt concentration is increased to Li$^+$: H$_2$O = 1:4 or even higher, some commonalities come into being. First, the heights of each peak do not change obviously. The following sequence is organized from high peaks to low ones: Li-Oc peak, H-Oc peak, Oc-Hc peak, O-Co peak and Li-C$_H$ peak. The rest RDF peaks are indistinct. Second, the O-Co peak becomes higher than the Li-C$_H$ peak, which does not occur in the dilute solutions. This may be caused by the fact that the interaction between Li$^+$ ion and CH$_3$COO$^-$ anion is strong enough and when the number of CH$_3$COO$^-$ anion increases, CH$_3$COO$^-$ anion tends to interact with the water molecules in the concentrated solutions. Although the RDF does not change obviously when the concentration reaches Li$^+$: H$_2$O = 1:4, there are still some small changes. For instance, the peak of Oc-Hc interaction is slightly weakened, indicating that the interaction between the CH$_3$COO$^-$ anion and the other CH$_3$COO$^-$ anion no longer increases, and, both the interaction between the CH$_3$COO$^-$ anion and the water molecule and the one between the CH$_3$COO$^-$ anion and lithium ion tend to increase more.

The variations of the concentrations are discussed in four conditions, respectively: 1:32 represents the most dilute condition, 1:8 the relatively more dilute one, 1:5.3 the relatively more concentrated one, and 1:4 and above the most concentrated one. It can be observed that when the concentration of the solution system is increased to $Li^+: H_2O = 1:4$, the 8 main RDF curves of the whole system change slightly, suggesting that the state of the solution system hardly changes. However, when the concentration is lower than $Li^+: H_2O=1:4$, the RDF curves of the whole system change obviously.

**Variations of interactions concerning anions.** **Fig. 3** indicates the variation of the main interactions related with $Li^+$ ions in the solutions from dilution to concentration. In **Fig. 3a**, in the RDF curves of $Li^+$-$Li^+$, the most scattered points appear in the dilute solutions. Due to the insufficient statistics caused by the insufficient number of $CH_3COOLi$ in the system, a continuous curve can by no means be obtained, though by comparison we can figure out that the first peak is near 4.4 Å. As indicated in **Fig. 3b**, the first peak of the RDF curves of Li-Oc is at 2.9 Å, and as the concentration increases, the height is continuously decreasing and tends to stabilize at the concentration of 1:4. The second peak is at 4.8 Å and tends to be stable at the concentration of 1:4, but it is only about half the height of the peak in a dilute solution. Oc is an oxygen atom on the $CH_3COO^-$ anion, whose first RDF peak with $Li^+$ ion is very high and second peak is only as 1/4 high as the first peak. This indicates that an O atom in the $CH_3COO^-$ anion has a stable interaction with the lithium ion at 2.9 Å and another O atom in the same $CH_3COO^-$ anion does not have an obvious interaction with the lithium ion. As a further result we can conclude that there is a fixed spatial configuration constituted by $Li^+$ ions and the Oc atom in their nearest $CH_3COO^-$ anion, but there is no fixed spatial configuration constituted by $Li^+$ ions and their nearest $CH_3COO^-$ anion. A comparison can thus be conducted between this noteworthy point and the RDF curve of Li-$C_H$ below. In **Fig. 3c**, it is evident that as the concentration increases, the variations of the whole RDF curves of Li-$C_H$ become more and more obvious. Besides, compared with other curves, the contrast in height between the first peak and the second peak is the smallest. Behind the second peak, the other peaks are indistinct. The first peak is at 3.5 Å, the second 4.8 Å. This demonstrates a clear formation of a triangle relation between $Li^+$ ion, $C_H$ atom and another $C_H$ atom.

It can be observed in **Fig. 4** that throughout the course of changing the solutions from dilution to concentration, the oxygen atom in the water molecule is studied, and it may have strong interaction with $C_O$ and $H_C$. The RDF curves of the O-Co interaction can be observed in **Fig. 4a**. Besides, it is true that there will be an obvious peak near 3.4 Å, suggesting that Co atom, a relatively inner atom in $CH_3COO^-$ connected with two O atoms and one C atom, still interacts strongly with O atom in water molecule at the same time. Another noteworthy point is that regardless of the concentrations, the variation of the RDF curves is not obvious. This can be explained as: on the one hand, because the number of water molecules is very large, there are enough water molecules around each anion to form a connection with it; on the other hand, there is no obvious variation of this local structure in different concentrated solutions. What is interesting about **Fig. 4b** is that there is no significant peak between H atom in water molecule and $C_H$ atom in $CH_3COO^-$ anion and the distribution of the whole space is relatively average.

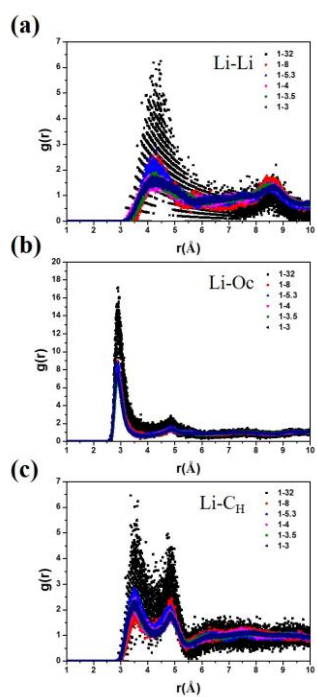

**Fig. 3.** The RDF curves of Li-Li, Li-$O_C$, Li-$C_H$ are compared respectively in $CH_3COOLi$ aqueous solutions with concentrations of $Li^+:H_2O$ = 1:32, 1:8, 1:5.3, 1:4, 1:3.5 and 1:3.

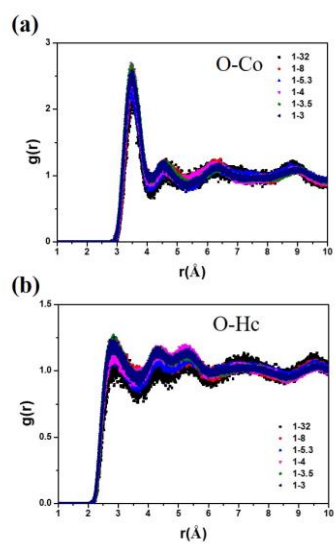

**Fig. 4.** The RDF curves of O-$C_O$, O-$H_C$ are compared respectively in $CH_3COOLi$ aqueous solutions with concentrations of $Li^+:H_2O$ = 1:32, 1:8, 1:5.3, 1:4, 1:3.5 and 1:3..

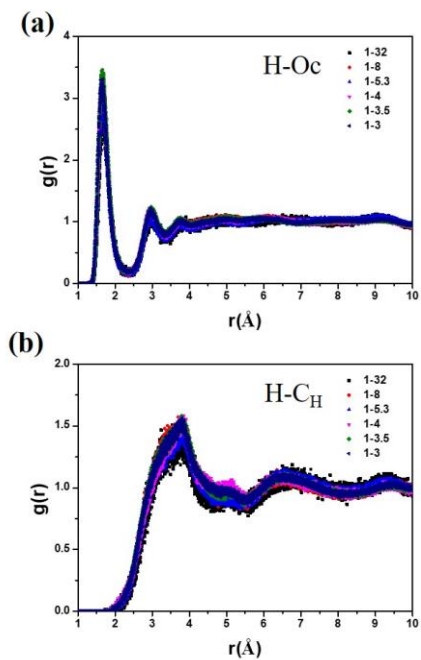

**Fig. 5.** The RDF curves of H-$O_C$, H-$C_H$ are compared respectively in $CH_3COOLi$ aqueous solutions with concentrations of $Li^+:H_2O$ = 1:32, 1:8, 1:5.3, 1:4, 1:3.5 and 1:3..

Next research object is H atom in the water molecule (**Fig. 5**). H atom has strong interactions with Oc atom and $C_H$ atom. In the RDF curves of H-Oc, the first peak is very obvious, at 1.7 Å, which is much higher than the second peak at 3 Å. This indicates that, on the one hand, there is a significant hydrogen bond between the H atom in the water molecule and the O atoms in $CH_3COO^-$ anion; on the other hand, there are no other obvious spatial configurations formed except the hydrogen bond. In the RDF curves of H-$C_H$, the first peak is highest but only of 1.5 in probability density. That is to say, the probability density of the appearance of H atom in a water molecule next to $C_H$ atom at 3.8 Å is only 50% higher than the average probability density of the whole space. It thus can be considered as a very weak interaction. Another point worth noticing is that the RDF curves of H-$C_H$ do not distinctly vary in different concentrations. At the end of this section, the RDF curves of Oc-Hc are brought into discussion. As the solution changes from dilute state to concentrated one, the whole curve falls down and only two peaks appear (**Fig. 6**). This indicates that there is only a short interaction between a $CH_3COO^-$ anion and another $CH_3COO^-$ anion. **Fig. 7**, as a schematic diagram, demonstrates the main characteristics of the variations of the microstructure in dilute and concentrated solutions. The distance between $Li^+$ ion and the $O_C$ on its nearest $CH_3COO^-$ anion remains 2.9 Å. In addition, the distance between $Li^+$ ion and the $C_H$ on its nearest $CH_3COO^-$ anion remains 3.5 Å, which remains constant in both dilute and concentrated solutions. But in dilute solutions, pairs of clusters are randomly scattered in space and are not restricted with each other. By contrast, in concentrated solutions, pairs of clusters construct a huge spatial structure, which may provide insurance for the electrochemical stability of the solution.

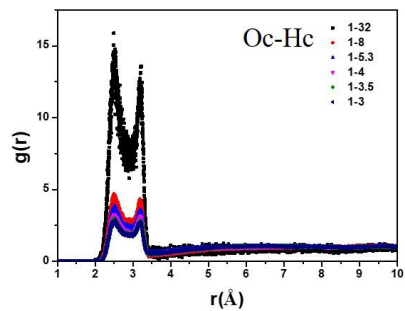

**Fig. 6.** The RDF curves of O-$H_C$ are compared respectively in $CH_3COOLi$ aqueous solutions with concentrations of $Li^+:H_2O$ = 1:32, 1:8, 1:5.3, 1:4, 1:3.5 and 1:3..

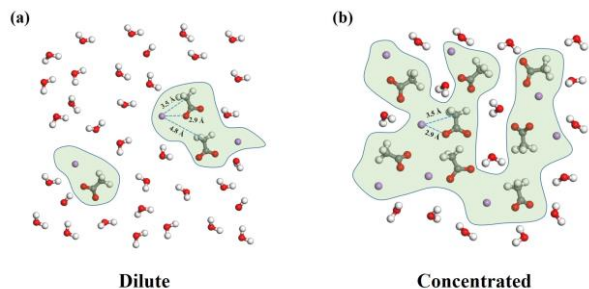

**Fig. 7.** Illustration of the evolution of the $Li^+$ primary solvation sheath from diluted to highly concentrated $LiNO_3$ aqueous solutions..

In summary, $CH_3COOLi$ is considered as a candidate as salute in aqueous electrolyte. The characteristics of the microstructure of the $CH_3COOLi$ aqueous system in different concentrations are discussed in detail, and anions are especially focused on. The RDF curves have obvious stability in the concentrated solutions. There is a fixed spatial configuration between Li$^+$ ions and the O$_c$ atom in their nearest CH$_3$COO$^-$ anion, but there is no fixed spatial configuration between Li$^+$ ions and their nearest CH$_3$COO$^-$ anion. A clear formation of a triangle relation between Li$^+$ ion, C$_H$ atom and another C$_H$ atom is discovered. Based on information provided by RDF, the largest variation throughout the change of concentration is the probability of the occurrence of each RDF peak when the RDF peaks are constants. This demonstrates that when the concentration changes, the microstructure of small components formed by any two clusters does not change much, but at same time, the spatial aggregation constructed by many small components is gradually built up from a broader perspective.

**Acknowledgement**

This work was financially supported by National Materials Genome Project (2016YFB0700600), the National Natural Science Foundation of China (No. 21603007 and 51672012), and Shenzhen Science and Technology Research Grant (No. JCYJ20150729111733470 and JCYJ20151015162256516).